\documentclass[12pt]{article} 
\usepackage{epsfig} 
\usepackage{amsfonts} 
\usepackage{amssymb}
\usepackage{amsmath}  
\usepackage{amsbsy}  
\usepackage{wasysym}
\usepackage{cite}
\usepackage{multirow} 
\input axodraw.sty
\catcode`\@=11 
\@addtoreset{equation}{section} 
\def\theequation{\thesection.\arabic{equation}} 
\@twosidetrue 
\catcode`\@=11  
\def\section{\@startsection{section}{1}{\z@}
{3.5ex plus 1ex minus .2ex}{2.3ex plus .2ex}{\large\bf}}
\def\thesection{\arabic{section}} 
\def\thesubsection{\arabic{section}.\arabic{subsection}} 
\def\thesubsubsection{
\arabic{section}.
\arabic{subsection}.
\arabic{subsubsection}}
\def\appendix{\setcounter{section}{0} 
\def\thesection{\Alph{section}} 
\def\theequation{\Alph{section}.\arabic{equation}} 
\def\thesubsection{\Alph{section}.\arabic{subsection}} 
\def\thesubsubsection{
\Alph{section}.
\arabic{subsection}.
\arabic{subsubsection}} 
\def\section{\@startsection{section}{1}{\z@}
{3.5ex plus 1ex minus .2ex}{2.3ex plus .2ex}{\large\bf}}} 
\newcount\minutes 
\newcount\scratch 
\def\timestamp{%
\scratch=\time 
\divide\scratch by 60 
\edef\hours{\the\scratch} 
\multiply\scratch by 60 
\minutes=\time 
\advance\minutes by -\scratch 
---$\,$\hours:\null 
\ifnum\minutes< 10 0\fi 
\the\minutes} 

\def\sla#1{\ifmmode%
\setbox0=\hbox{$#1$}%
\setbox1=\hbox to\wd0{\hss$/$\hss}\else%
\setbox0=\hbox{#1}%
\setbox1=\hbox to\wd0{\hss/\hss}\fi%
#1\hskip-\wd0\box1 } 
\def\waa{{$W^\pm\gamma\gamma$}}

\def\vbfnlo{{\tt VBFNLO}}

\def\beq{\begin{equation}} 
\def\eeq{\end{equation}} 
 
\def\beqn{\begin{eqnarray}} 
\def\eeqn{\end{eqnarray}}

\def\({\left(} 
\def\){\right)} 
\def\as{\ifmmode \alpha_s \else $\alpha_s$ \fi}

\begin{document} 
\begin{titlepage} 
\nopagebreak 
{\flushright{ 
    \begin{minipage}{5cm}
      FTUV-11-0324 \\
      IFUM-971-FT\\
      KA-TP-08--2011  \\  
      LPN11-14       \\  
      SFB/CPP-11-14 \\
      \end{minipage}
  } 
} 
\vfill 
\begin{center} 
{\LARGE \bf 
 \baselineskip 0.5cm 
\waa~production with leptonic decays at NLO QCD} 
\vskip 0.5cm  
{\large   
  G. Bozzi$^1$, F. Campanario$^{2}$, M. Rauch$^{2}$ and D. Zeppenfeld$^2$
}   
\vskip .2cm
{$^1$ {\it Dipartimento di Fisica, Universit\`a di Milano and INFN,
    Sezione di Milano\\ Via Celoria 16, I-20133 Milano, Italy}}\\
{$^2$ {\it Institut f\"ur Theoretische Physik, 
    Universit\"at Karlsruhe, KIT,\\ 76128 Karlsruhe, 
    Germany}}\\
\vskip 1.3cm     
\end{center} 
\nopagebreak 
\begin{abstract}
The computation of the ${\cal O}(\alpha_s)$ QCD corrections to the cross
sections for $W^\pm \gamma \gamma$ production in hadronic collisions is
presented. We consider the case of real photons in the final state, but
include full leptonic decays of the $W$. Numerical results for the LHC
and the Tevatron are obtained through a parton level
Monte Carlo based on the structure of the VBFNLO program, allowing an
easy implementation of general cuts and distributions. We show the
dependence on scale variations of the integrated cross sections and
provide evidence of the fact that NLO QCD corrections strongly modify
the LO predictions for observables at the LHC both in magnitude and in
shape.
\end{abstract} 
\vfill 
\today \timestamp 
\hfill 
\vfill 

\end{titlepage} 

\newpage               

\section{Introduction}
\label{sec:intro}
Precise and reliable predictions of cross sections at hadron colliders require
the calculation of higher order QCD corrections. As part of such a program we
have in the past determined next-to-leading-order (NLO) QCD corrections for
the production cross sections of various combinations of three electroweak
bosons, including $W^+W^-Z$~\cite{Hankele:2007sb}, $W^\pm W^\mp W^\pm$ and 
$W^\pm ZZ$~\cite{Campanario:2008yg}, $W^+W^-\gamma (ZZ\gamma)$~\cite{last} and 
$W^\pm Z\gamma$ production~\cite{Bozzi:2010sj}. In all cases, leptonic
decays of the weak bosons were included in the calculations. For the
production of three weak bosons, these results were verified against
independent calculations~\cite{Lazopoulos:2007ix,Binoth:2008kt} which
are available for on-shell weak bosons and neglecting Higgs boson
exchange.  

In the present paper, we present the ${\cal  O}(\alpha_s)$ QCD 
corrections for processes with two photons and one $W$
in the final state, namely the production of 
\begin{eqnarray}
\label{processes}
"W^+\gamma\gamma" \qquad & pp,\; p\bar p \to 
\nu_{l} l^+ \gamma \gamma~+X  \nonumber \\ 
"W^-\gamma \gamma" \qquad & pp,\; p\bar p \to l^-\bar{\nu}_l \gamma
\gamma~+X \, .
\end{eqnarray}
A calculation of $W^+\gamma\gamma$ production at NLO QCD accuracy has been performed
before~\cite{Baur:2010zf}, including the $q \to q \gamma$ fragmentation
contributions. However, also Ref.~\cite{Baur:2010zf} treats the $W$ as
stable. By including off shell effects and, in particular, photon radiation from
the final state charged lepton we aim at providing complementary information.

Processes involving two or three electroweak bosons in the final
state are relevant for studying anomalous gauge interactions, as they give
direct access to triple and quartic couplings~\cite{quartic}. \waa~production
is sensitive to the $W W \gamma$ and $WW\gamma\gamma$  vertices. In addition,
a final state with two photons and missing transverse energy is relevant in a
variety of beyond the standard model scenarios~\cite{Campbell:2006wx}: in
gauge-mediated supersymmetry breaking, for instance, the neutralino is
often the next-to-lightest supersymmetric particle and decays into a
photon plus a gravitino, giving a signal of two photons and missing $E_T$. In
Ref.~\cite{CMS2009}, a study of the background estimates for
supersymmetry motivated di-photon production searches has been performed,
pointing out the relevance of the \waa~production process as a standard
model~(SM) background in case of electron misidentification. Another possible
application is an estimate of backgrounds when searching for $WH$ production,
followed by Higgs decay to two photons. 

The present work closely follows our previous calculations of NLO QCD
corrections to triple electroweak boson production. In particular,
photon isolation is implemented within the Frixione
approach~\cite{Frixione:1998jh} and we, thus, avoid the need for the
inclusion of $q \to  q \gamma$ fragmentation contributions, which
were discussed for $W^+\gamma\gamma$ production by Baur et al.~\cite{Baur:2010zf}. 
Similar to previous work on triple weak boson production, we find that
the QCD corrections are sizable and also modify the shape of the
differential distributions for many observables: this proves that a
simple rescaling of the LO results is not adequate and a full NLO Monte
Carlo is needed for any quantitative determination of quartic couplings
at the LHC. We have implemented our calculation within the VBFNLO
framework~\cite{Arnold:2008rz}, a parton level Monte
Carlo program which allows the definition of general acceptance cuts
and distributions.

The paper is organized as follows: in Section~\ref{sec:calc} we show the
Feynman diagrams relevant for the calculation and provide an overview of
the strategies used to compute the real and virtual corrections and the
various checks performed to ensure the numerical accuracy of our
code. In Section~\ref{sec:res}, we show numerical results, including the
scale variations of the LO and NLO integrated cross sections and some
selected differential distributions. Particular concern will be given to
the consequences of approximate radiation zeroes which largely disappear when
going from LO to NLO distributions, as was already emphasized in
Ref.~\cite{Baur:2010zf}. 
Conclusions are given in Section~\ref{sec:concl}.

\section{The calculation}
\label{sec:calc}
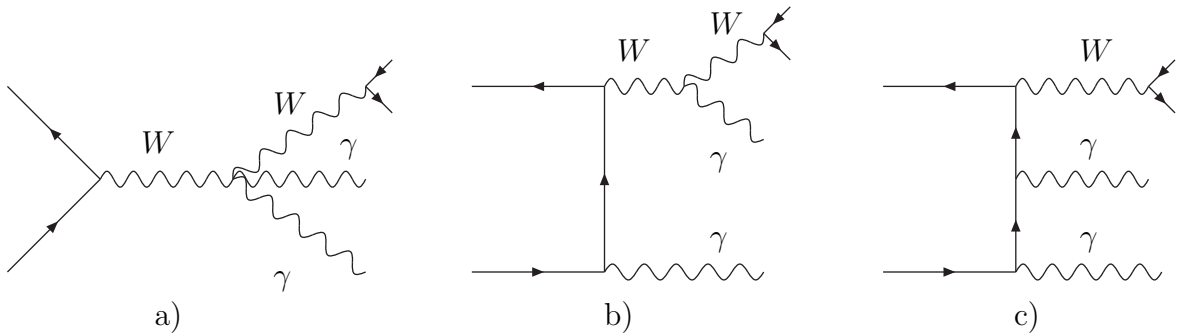
\begin{figure}[hbtp]
  \begin{center}
    \setlength{\unitlength}{1pt}
    \begin{picture}(50,110)(0,0)
      \ArrowLine(-195,15)(-160,50)
      \ArrowLine(-160,50)(-195,85)
      \Photon(-160,50)(-110,50){3}{5}
      \Photon(-110,50)(-60,85){3}{5}
      \Photon(-110,50)(-60,50){3}{5}
      \Photon(-110,50)(-60,15){3}{5}
      \ArrowLine(-50,95)(-60,85)
      \ArrowLine(-60,85)(-50,75)
      \put(-145,60){$W$}
      \put(-95,75){$W$}
      \put(-70,60){$\gamma$}
      \put(-95,10){$\gamma$}
      \put(-140,-5){a)}
      
      \ArrowLine(-20,15)(30,15)
      \ArrowLine(30,85)(-20,85)
      \ArrowLine(30,15)(30,85)
      \Photon(30,15)(90,15){3}{5}
      \Photon(30,85)(60,85){3}{3}
      \Photon(60,85)(90,105){3}{3}
      \Photon(60,85)(90,65){3}{3}
      \ArrowLine(90,105)(100,95)
      \ArrowLine(100,115)(90,105)
      \put(35,95){$W$}
      \put(70,105){$W$}
      \put(70,55){$\gamma$}
      \put(70,25){$\gamma$}
      \put(30,-5){b)}
      
      \ArrowLine(135,15)(185,15)
      \ArrowLine(185,85)(135,85)
      \ArrowLine(185,15)(185,50)
      \ArrowLine(185,50)(185,85)
      \Photon(185,85)(235,85){3}{5}
      \Photon(185,50)(235,50){3}{5}
      \Photon(185,15)(240,15){3}{5}
      \ArrowLine(245,95)(235,85)
      \ArrowLine(235,85)(245,75)
      \put(210,95){$W$}
      \put(210,60){$\gamma$}
      \put(210,25){$\gamma$}
      \put(185,-5){c)}
    \end{picture}  \\
    \setlength{\unitlength}{1pt}
    \caption{Examples of the three topologies of Feynman diagrams
      contributing to the process $pp\to$\waa + X \, at tree-level.}
    \label{fig:1}
  \end{center}
\end{figure}

For the calculation of virtual corrections, it is convenient to classify
the contributing Feynman graphs according to the number of electroweak
boson vertexes which are attached to the quark lines: graphs with three
vertexes lead to pentagon loops, while boxes and triangles are the most
complex structures for graphs with two or one electroweak vertex on the
quark line. Examples for the three classes of diagrams contributing to
\waa~production at Born level are shown in Fig.~\ref{fig:1}.  
Diagrams (a) and (b) exhibit
the dependence on the quartic ($WW\gamma\gamma$) and triple ($WW\gamma$)
gauge boson couplings. 

As is customary in all VBFNLO calculations, we
include the full spin correlations of the leptons coming from the $W$
decay and also the final state photon radiation off the charged
leptons. As in all previous calculations, we have factorized the
subgraphs involving boson splitting and decay into {\it leptonic
  tensors}~\cite{Jager:2006zc}, computed once per phase-space point in
order to speed up the code, and we made use of the helicity technique
introduced in Ref.~\cite{Hagiwara:1985yu} for the computation of matrix
elements.
The cancellation of infrared divergences coming from real and virtual
corrections at NLO is achieved using the Catani-Seymour dipole
subtraction method~\cite{Catani:1996vz}. We note here that the strategy
of separately computing the {\it leptonic tensors} proves particularly
useful given the large number (372) of real emission diagrams at NLO.

For the real emission contributions, there is an additional issue to consider:
the infrared singularity that may result from a photon emitted
collinearly to a massless quark. The simple rejection of any event
containing partons within a cone drawn around the photon direction would
inevitably spoil the cancellation of infrared divergences for
subprocesses with a gluon in the final state. As a solution, we use
the {\it photon isolation cut} proposed by Frixione~\cite{Frixione:1998jh} which is described in more details in
Section~\ref{sec:res}.

The NLO virtual corrections result from one-loop diagrams obtained by
attaching a gluon line to the quark-antiquark line in diagrams like the
ones depicted in
Fig.~\ref{fig:1}. We combine the virtual corrections into three
different groups, which  include all loop diagrams derived from a given
Born level configuration~\cite{Jager:2006zc}. This leaves us with three
universal building 
blocks, namely corrections to one, two or three vector bosons attached
to the quark line. In all cases, the infrared divergent contributions
factorize in terms of their corresponding Born amplitude and the full
virtual term is given by
\begin{align}
  M_V = \widetilde{M}_V + \ \frac{\alpha_S}{4 \pi} \ C_F \ \left( \frac{4
      \pi \mu^2}{Q^2} \right)^\epsilon  \ \Gamma{(1 + \epsilon)} \ \left[
    -\frac{2}{\epsilon^2} - \frac{3}{\epsilon} - 8 + \frac{4 \ \pi^2}{3}
  \right] \ M_B, \label{eq:MV}
\end{align}
where $Q$ is the partonic center-of-mass energy, i.e. the invariant mass
of the \waa~system,  $M_B$ the total born contribution, and the term
$\widetilde M_V$  includes the finite parts of the virtual corrections
to 2 and 3 weak boson amplitudes which we call virtual-box and
virtual-pentagons in the following, so named since boxes and pentagons
constitute the most complex loop diagrams, respectively. 
These finite terms can be calculated in $d=4$ dimensions. The tensor
coefficients of the loop integrals have 
been computed by means of the Passarino-Veltman reduction formalism up
to the box level, but avoiding the explicit calculation of inverse 
Gram matrices by solving a system of linear equations which is a more
stable procedure close to singular points. For pentagons, we use the
Denner-Dittmaier reduction formalism~\cite{Denner:2002ii},  which is now
fully implemented in the public version of the \vbfnlo~code for all
multi-boson processes. As in our 
previous calculation~\cite{last}, we explicitly checked that there are no
additional infrared singularities other than those proportional to the
Born amplitude. 

A powerful test of the virtual corrections is provided by Ward
identities which connect virtual-pentagon and virtual-box contributions.
For this purpose we write
the polarization vector of the $W$ as~\cite{Jager:2006zc}
\begin{align}
  \label{mapping}
  \epsilon_{W}^\mu  = x_{W} \, q_{W}^\mu + \tilde{\epsilon}_{W}^\mu,
\end{align}
where $q_W$ is the momentum of the $W$ and the remainder
$\tilde{\epsilon}_{W}^\mu$ is chosen in such a way that 
\begin{align}
  \tilde{\epsilon}_{W} \cdot (q_{\gamma_1} + q_{\gamma_2}) = 0 \, ,
\end{align}
i.e., the time component of the shifted polarization vector is zero in
the center-of-mass system of the photon pair. Via  Ward identities, a
pentagon contracted with 
the $W$ momentum, $q_W$, can always be reduced
to a difference of boxes: in this way we ''shift'' a fraction of
pentagon diagrams to box subroutines. The fact that the sum of the
virtual contributions does not change upon this shift 
provides a very powerful consistency check of our implementation.

The numerical accuracy of our code for tree level amplitudes has been
tested against {\tt MadGraph}~\cite{Stelzer:1994ta} at the level of
amplitudes and against {\tt Sherpa}~\cite{sherpa}  for integrated cross
sections, finding agreement at the level of machine precision and  at the
per mill level, respectively. 
As a final test, we have made a comparison with the numbers of the
proceeding paper of Ref.~\cite{Baur:2010zf} for the production of
on-shell bosons without leptonic decays. For that we have neglected all
diagrams with the photon emitted from the lepton line and used
narrow-width approximation for the vector boson decay. In
Table~\ref{Comparison}, we show the comparison for $W^+\gamma \gamma +X$ production
at LO at the LHC. We do not present an explicit comparison at NLO since we do
not include fragmentation functions and our isolation procedure for the
photon differs from the one used in Ref.~\cite{Baur:2010zf}. As input
parameters for this comparison, we use $\alpha$ at $ q^2=0$ and standard
FormCalc EW parameters. 
\begin{table}[ht!]
  \begin{center}
    \renewcommand{\arraystretch}{1.15}
    \begin{tabular*}{0.77\textwidth}{@{\extracolsep{\fill}}|c@{\hspace{0.6cm}}|c@{\hspace{1.8cm}}|c@{\hspace{1.4cm}}|}
      \hline
      Scale &  Program   &   LO [fb] \\ \hline
      \multirow{2}{*}{100 GeV} &
      VBFNLO &   7.317(1)   \\
             &Ref.~\cite{Baur:2010zf} &7.253(5) \\
         \hline
    \end{tabular*}
    \caption[]{Comparison between our results and the ones of Ref~.\cite{Baur:2010zf}
      for $pp \to  W^+ \gamma \gamma $  $+X$ at LO at the LHC. The input parameters and
      settings are taken from Ref.~\cite{Baur:2010zf}.}
    \label{Comparison}
  \end{center}
\end{table}

Additionally, to
control the  numerical stability of our code, a gauge test based on Ward identities has
been used throughout the virtual implementation for each phase space point. 

\section{Results}
\label{sec:res}
\subsection{Definition of scales and cuts}

We have implemented our calculation as a NLO Monte Carlo program based
on the structure of the \vbfnlo \, code~\cite{Arnold:2008rz}. 
For the electroweak parameters, we use the $W$ and $Z$ boson masses and 
the Fermi constant as input. From these, we derive the electromagnetic 
coupling and the weak mixing angle via tree level relations, i.e. we use 
\begin{align}
  &m_W = 80.398 \ \mathrm{GeV}  & m_Z = 91.1876 \ \mathrm{GeV} &~~~~~~~~\sin^2{(\theta_W)} = 0.22264 \nonumber\\
  &G_F = 1.16637 \cdot 10^{-5} \ \mathrm{GeV}^{-2} 
  &\alpha^{-1} = 132.3407.~~~ \; & & \label{eq:ew}
\end{align}
We do not consider bottom and top quark effects. The remaining quarks are
assumed to be massless and we work in the approximation where 
the CKM matrix is the identity matrix.
We choose the invariant ``\waa'' mass as the central value for the
factorization and renormalization scales: 
\begin{align} \label{eq:WAAmass}
  \mu_F = \mu_R = \mu_0 = \sqrt{(p_{\ell} + p_{\nu} 
    + p_{\gamma_1} + p_{\gamma_2})^2}.
\end{align}
We use the CTEQ6L1~\cite{Pumplin:2002vw} parton distribution function at
LO and the CT10~\cite{Lai:2010vv}
set with $\alpha_S(m_Z)=0.1180$ at NLO.

A real photon in the final state can be emitted either from the initial
quark line or from the final-state charged lepton. For efficient Monte
Carlo generation, we divide the phase space into three separate regions 
to consider all the possibilities and then sum the contributions to get the total
result. The regions are generated as triple electroweak boson production
as well as $W\gamma_1$ and $W\gamma_2$ production with (approximately) on-shell $W^+\to
l^+\nu_l\gamma$ (or  $W^-\to \l^-\bar\nu_l\gamma$) three-body decay, respectively. 

We impose a set of minimal cuts on the rapidity, $y_{\ell(\gamma)}$, and
the transverse momenta, $p_{T{\ell(\gamma)}}$, of the charged leptons
and the photon, which are designed to represent typical experimental 
requirements. Furthermore, leptons, photons and jets must be well
separated in the rapidity-azimuthal angle plane. Specifically, the cuts
imposed are
\begin{equation}
  p_{T{\ell(\gamma)}} > 20 \ \mathrm{GeV} \qquad
  |y_{\ell(\gamma)}| < 2.5 \qquad
  R_{\gamma\gamma} > 0.4  \qquad
  R_{\ell\gamma} > 0.4  \qquad
  R_{j \ell} > 0.4  \qquad
  R_{ j \gamma} > 0.7 
  \label{eq:cuts}
\end{equation}
where, in our simulations, a jet is defined as a colored parton of 
transverse momentum $p_{T j}>30$ GeV and rapidity $|y_{j}|<4.5$.

Since we do not include any fragmentation contribution, we must find a
way to reject events in which a quark and a photon are collinear in the
final state, i.e. we have to provide a prescription to {\it isolate} the
photons. We choose to implement the procedure defined in 
Ref.~\cite{Frixione:1998jh}: if $i$ is a parton with transverse energy
$E_{T_i}$ and has a separation $R_{i\gamma}$ with a photon of transverse
momentum $p_{T\gamma}$, then the event is accepted only if
\beq
\Sigma_i \, E_{T_i} \, \theta (\delta - R_{i\gamma}) \, \leq \,
p_{T\gamma} \, \frac{1-\cos\delta}{1-\cos\delta_0} \,\,\,\,\,\,\,\,\,\,
(\mathrm{for\,all} \,\,\,\,\,\delta\leq\delta_0) \label{eq:isol}
\eeq
where $\delta_0$ is a fixed separation which we set equal to 0.7. A quick
look at Eq.~(\ref{eq:isol}) reveals that a sufficiently soft parton can be 
arbitrarily close to the photon axis, while the energy of an exactly
collinear parton must be vanishing in order to pass the isolation
cut. Collinear-only events (leading to fragmentation contributions) are
thus rejected while soft emissions are retained as desired. 

\subsection{Integrated results}
In Table~\ref{LHCnumbers}, we give results for the integrated cross 
sections for \waa~production at the LHC (14 TeV) for the given cuts as
well as for a harder cut on the photon transverse momentum of
$p_{T\gamma}>30 $ GeV: the NLO corrections are large, enhancing the LO
result by more than a factor of 3 in all cases as can be seen by the
K-factor defined as K$=\sigma^{NLO}/\sigma^{LO}$.  This value is
larger than the ones in other triple vector boson production 
channels, where we observe K-factors between 1.5 and 2. However, it is
consistent with the results of Ref.~\cite{Baur:2010zf}, where a K-factor
of 2.93 was found after isolation cuts (which differ from ours). We
already note here that the large K-factor can be explained by the suppression
of the LO cross section due to the so called radiation zero. This will
be further investigated in Section~\ref{sec:Rapidis}. 

\begin{table}[ht!]
  \begin{center}
    \renewcommand{\arraystretch}{1.15}
    \begin{tabular*}{0.95\textwidth}{@{\extracolsep{\fill}}|c@{\hspace{0.6cm}}|c@{\hspace{0.8cm}}|c@{\hspace{0.8cm}}|c@{\hspace{0.5cm}}|}
      \hline
      LHC ($\sqrt{s}=14$~TeV) &  LO  [fb]     &   NLO [fb] & K-factor \\ \hline
      $~~\sigma("W^+\gamma\gamma" \to e^+ \nu_e \gamma \gamma)$ &&&\\
    $~~p_{T{\gamma(\ell)}}>20(20)$ GeV &
      \multirow{1}{*}{2.529}
      &\multirow{1}{*}{7.940}
       &3.14\\ 
     $~~p_{T{\gamma(\ell)}}>30(20)$ GeV
      &\multirow{1}{*}{ 0.979}
       &\multirow{1}{*}{3.172}
      &3.24\\
       \hline
      $~~\sigma("W^-\gamma\gamma" \to  e^- \bar \nu_e \gamma \gamma)$ &&&\\
       $~~p_{T{\gamma(\ell)}}>20(20)$ GeV
      &\multirow{1}{*}{1.946}
      &\multirow{1}{*}{6.759}
       &3.47\\ 
      $~~p_{T{\gamma(\ell)}}>30(20)$ GeV       
      &\multirow{1}{*}{0.686}
       &\multirow{1}{*}{2.583}
      &3.77\\
      \hline
    \end{tabular*}
    \caption[]{Total cross sections at the LHC for $pp \to $ \waa$+X$ with
      leptonic decays, at LO and NLO, and for two sets of cuts. Relative
      statistical errors of the Monte Carlo are below $10^{-3}$.}
    \label{LHCnumbers}
  \end{center}
\end{table}

In Table~\ref{Tevnumbers}, the numbers for the Tevatron (1.96 TeV) are
presented for the cuts as in Eq.~(\ref{eq:cuts}) (and for photon
isolation as in Eq.~(\ref{eq:isol})), but for
less restrictive transverse momentum cuts, $p_{T{\gamma(\ell)}}> 10(10)$ GeV
and $p_{T{\gamma(\ell)}}> 20(10)$ GeV. We only show the
$W^+\gamma\gamma$ case, because $W^-\gamma\gamma$ production is exactly
symmetrical at a $p\bar p$ collider. The NLO enhancement is less
pronounced for this collider, but still amounts to 60-70\% of the LO
result.

\begin{table}[ht!]
  \begin{center}
    \renewcommand{\arraystretch}{1.15}
    \begin{tabular*}{0.95\textwidth}{@{\extracolsep{\fill}}|c@{\hspace{0.6cm}}|c@{\hspace{0.8cm}}|c@{\hspace{0.8cm}}|c@{\hspace{0.5cm}}|}
      \hline
      Tevatron&  LO  [fb]     &   NLO [fb] & K-factor\\ \hline
      $~~\sigma("W^+\gamma\gamma" \to e^+ \nu_e \gamma \gamma)$ &&&\\
      $~~p_{T{\gamma(\ell)}}>10(10)$ GeV
      &\multirow{1}{*}{4.779}
      &\multirow{1}{*}{7.558}
      &1.58\\ 
      $~~p_{T{\gamma(\ell)}}>20(10)$ GeV
      &\multirow{1}{*}{0.5591}
      &\multirow{1}{*}{0.9415}
      & 1.68
      \\ \hline
    \end{tabular*}
    \caption[]{Total cross sections at the Tevatron for $p\bar{p} \to
      W^+\gamma\gamma+X$ with leptonic decays, at LO and NLO, and for two
      sets of cuts. Relative statistical errors of the Monte Carlo are
      below $10^{-3}$.}
    \label{Tevnumbers}
  \end{center}
\end{table}

In the following, we include a combinatorial factor of 2 in all figures
which corresponds to the production of electrons and muons. We have
studied the scale uncertainty of the total cross section by 
varying the renormalization and factorization scales as
\beq
\mu_F, \mu_R = \xi\cdot \mu_0 \,\,\,\,\,\,\,\,\,\,\,\,\, (0.1 < \xi < 10).
\label{eq:scale}
\eeq
\begin{figure}[tbh!]
  \includegraphics[scale = 1]{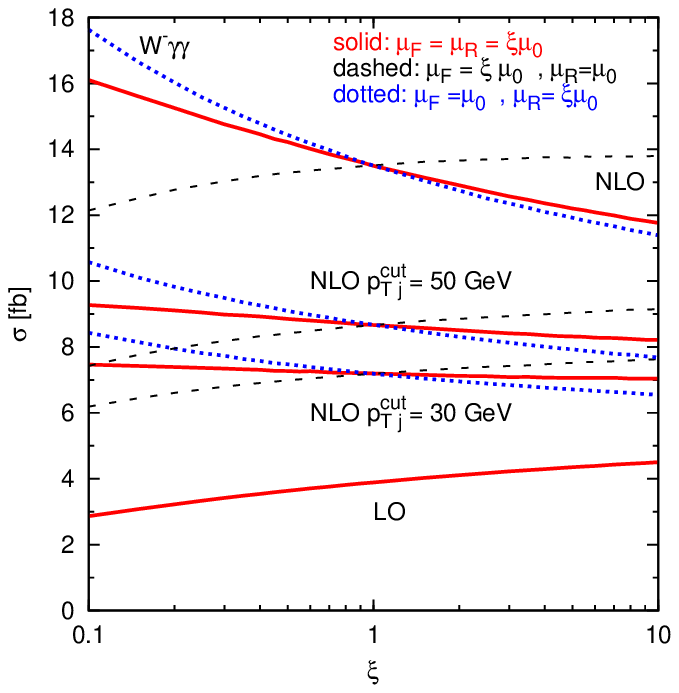}
  \includegraphics[scale = 1]{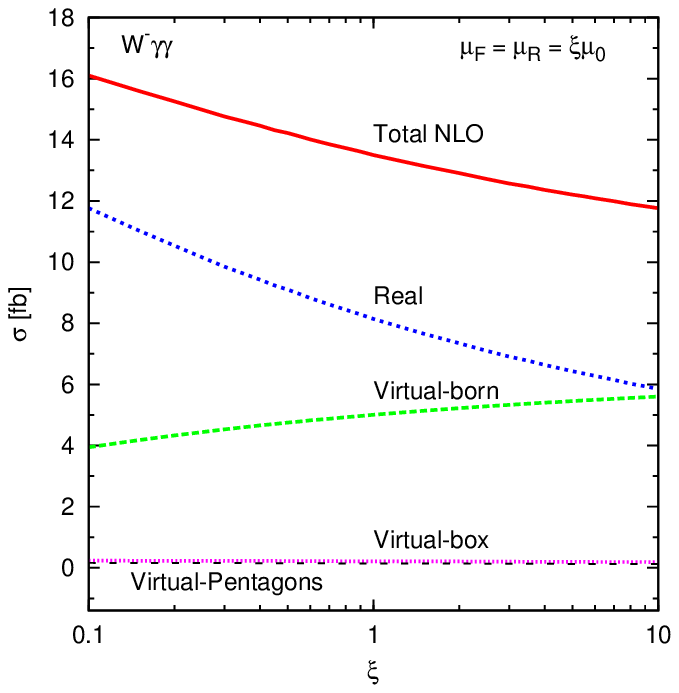}
  \caption[]{\label{fig:2}
    {\it Left:} {\sl Scale dependence of the total LHC cross section for 
      $p p \to W^-\gamma\gamma +X \to \ell^- \gamma \gamma +\sla{p}_T+X$ at
      LO and NLO within the cuts of Eqs.~(\ref{eq:cuts}, \ref{eq:isol}).
      The factorization and renormalization scales are together or
      independently varied in the range from $0.1 \cdot \mu_0$ to $10 \cdot
      \mu_0$.} 
    {\it Right:} {\sl Same as in the left panel but for the different NLO
      contributions at $\mu_F=\mu_R=\xi\mu_0$.}}
\end{figure}
In Fig.~\ref{fig:2} and Fig.~\ref{fig:3}, we show numerical results for
$W^-\gamma\gamma$ production and $W^+\gamma\gamma$ production,
respectively, within the cuts of Eqs.~(\ref{eq:cuts}, \ref{eq:isol}).
On the left panel of each figure, we show the overall scale variation of
our numerical predictions at LO and NLO: as already shown in
Table~\ref{LHCnumbers}, the NLO K-factor is large both in absolute value
($\sim$ 3) and compared to the LO scale variation. The NLO scale
uncertainty is about 10$\%$ when varying the factorization and the
renormalization scale $\mu=\mu_F=\mu_R$ up and down by a factor 2 around
the reference scale $\mu_0=m_{W\gamma\gamma}$ and is mainly driven by the dependence on
$\mu_R$, which gives a negative slope with increasing energy, while the
dependence on $\mu_F$ shows the opposite behaviour. In the left panels,
we also show results for additional jet veto cuts, requiring $p_{T j}<50$~GeV
or $p_{T j}<30$~GeV.
From these curves, we can see that a large contribution to
the total cross section is due to real jet radiation. While it
is evident that the renormalization scale variation is highly reduced by
a jet veto, this reduction should not be interpreted as a smaller
uncertainty of the vetoed cross section: a similar effect in $WZj$
production could be traced to cancellations between different regions of
phase space and, thus, the small variation is
cut-dependent~\cite{Campanario:2010hp}.

\begin{figure}[tbh!]
  \includegraphics[scale = 1]{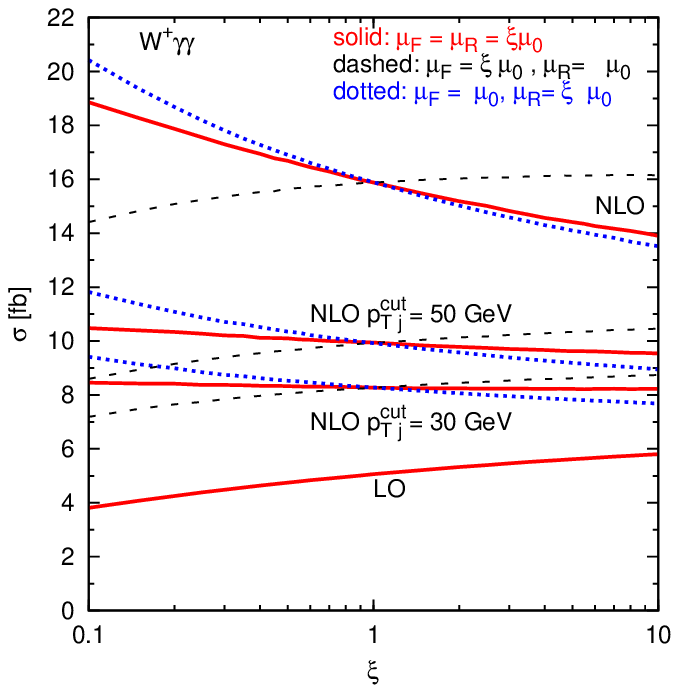}
  \includegraphics[scale = 1]{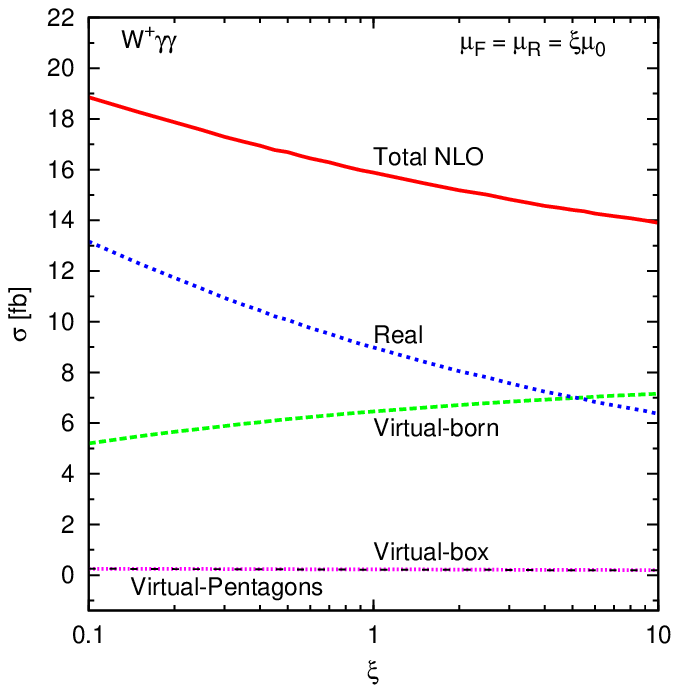}
  \caption[]{\label{fig:3}
     {\sl Same as Fig.~\ref{fig:2}, but for 
      $p p \to W^+\gamma\gamma +X \to \ell^+ \gamma \gamma +\sla{p}_T+X$ at
      the LHC.}}
\end{figure}

Since the 1-jet contributions to the ${\cal O} (\alpha_s)$ cross section
are determined at LO only, their scale variation is large. In fact, most
of the scale variation of the total NLO result is
accounted for by the real emission contributions, defined here as the
real emission cross section minus the Catani-Seymour subtraction terms
plus the finite collinear terms. This is more visible in the right
panels, where we show the scale dependence and compare the
size of the different parts of the NLO calculation.  As for the relative size of the NLO
terms, the real emission contributions dominate and are even larger than
the LO terms plus virtual terms proportional to the Born
amplitude. Non-trivial virtual contributions, namely the interference of
the Born amplitude with virtual-box and virtual-pentagon contributions,
represent less than 1\% of the total result and their scale 
dependence is basically flat.

\subsection{Differential cross sections}

Our numerical results show that the NLO corrections have a strong
dependence on the phase space region under investigation. As a
consequence, a simple rescaling of the LO results with a constant
K-factor is not allowed. As practical examples, we plot several
differential distributions at LO and NLO together with the associated
K-factor, defined as
\begin{align}\label{eq:kfactor}
  K = \frac{d \sigma^{NLO}/ dx}{d \sigma^{LO} /dx} \,\, ,
\end{align}
where $x$ denotes the considered observable. We usually show only one of
the $W^+\gamma \gamma$ and $W^- \gamma \gamma$ distributions in the
following since they share similar behaviors.  Also, we include in all
figures the results with an additional jet veto cut, requiring $p_{T j}<50$ GeV.

\begin{figure}[tb!]
  \includegraphics[scale = 1]{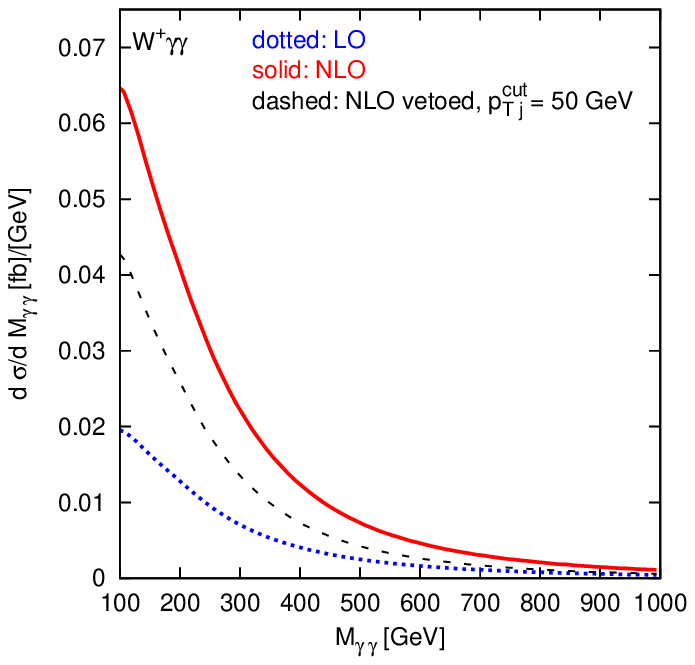}
  \includegraphics[scale = 1]{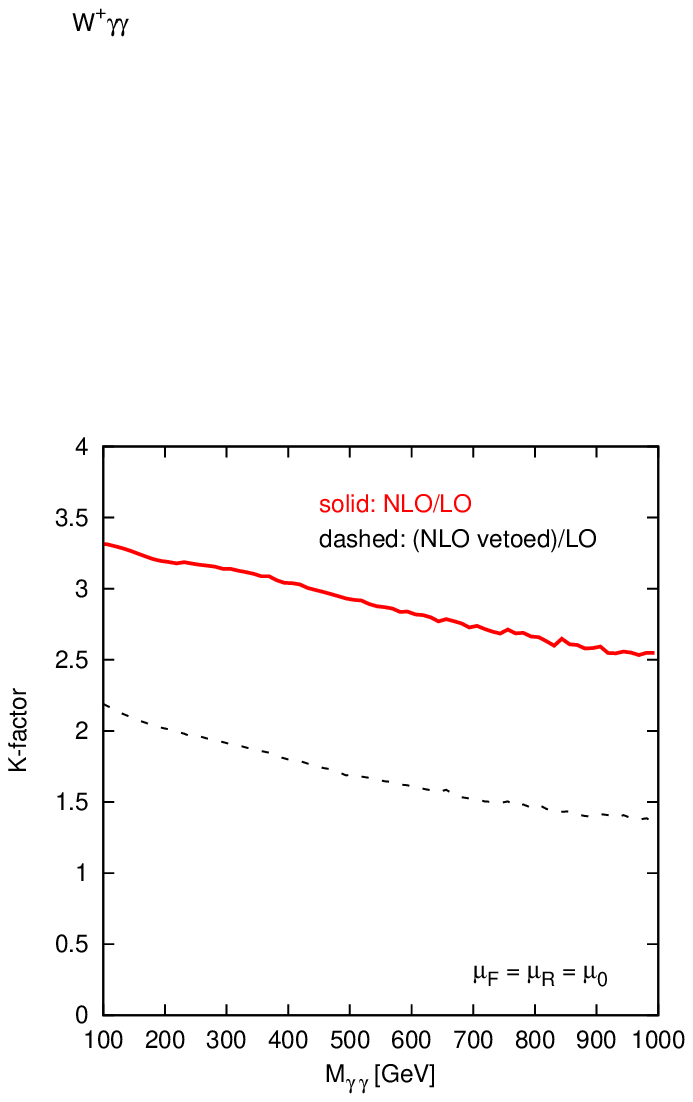}
  \caption[]{\label{fig:4}
    {\it Left:} {\sl Invariant mass distribution of the photon pair for
      $p p \to W^+ \gamma\gamma +X \to \ell^+ \gamma \gamma +\sla{p}_T+X$
      production at the LHC. LO and NLO results are shown for
      $\mu_F=\mu_R=\mu_0$ and the cuts of Eqs.~(\ref{eq:cuts}, \ref{eq:isol}).}
    {\it Right:} {\sl K-factor as defined in Eq.~(\ref{eq:kfactor}).}}
\end{figure}

In Fig.~\ref{fig:4}, we show the differential cross
section as a function of the invariant mass of the photon pair,
$M_{\gamma\gamma}$ for $W^+\gamma\gamma$ production. 
The corresponding K-factors lie between 2.5 and 3.5 in most
of the phase space region. 

\begin{figure}[tbh!]
 \includegraphics[scale = 1]{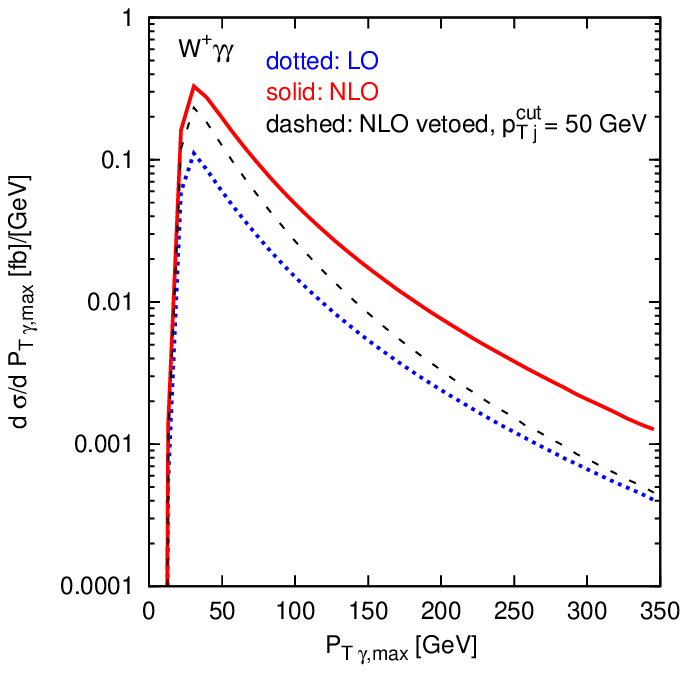}
  \includegraphics[scale = 1]{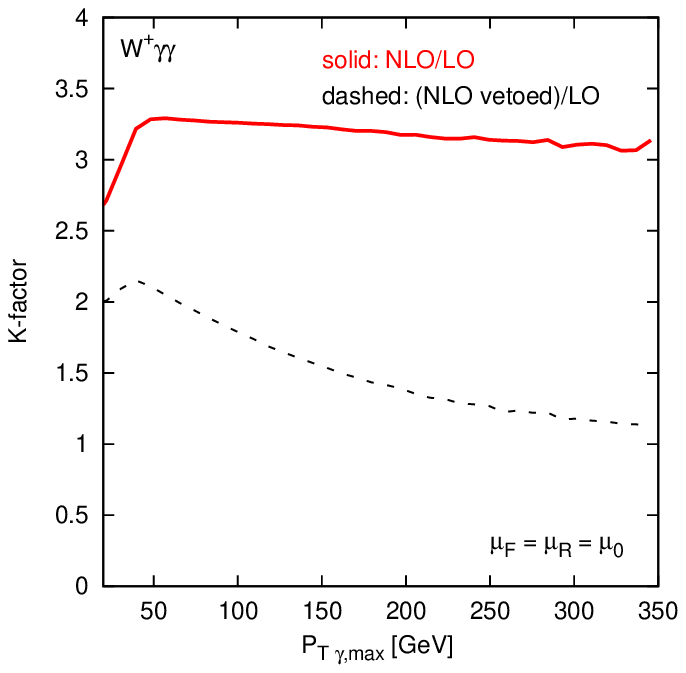}
  \caption[]{\label{fig:6}
    {\it Left:} {\sl Transverse-momentum distribution of the hardest
      photon in $p p \to W^+\gamma\gamma +X \to \ell^+ \gamma \gamma
      +\sla{p}_T+X$ production at the LHC. LO and NLO results are shown for
      $\mu_F=\mu_R=\mu_0$ and the cuts of Eqs.~(\ref{eq:cuts}, \ref{eq:isol}).} 
    {\it Right:} {\sl K-factor as defined in Eq.~(\ref{eq:kfactor}).}}
\end{figure}

\begin{figure}[tbh!]
  \includegraphics[scale = 1]{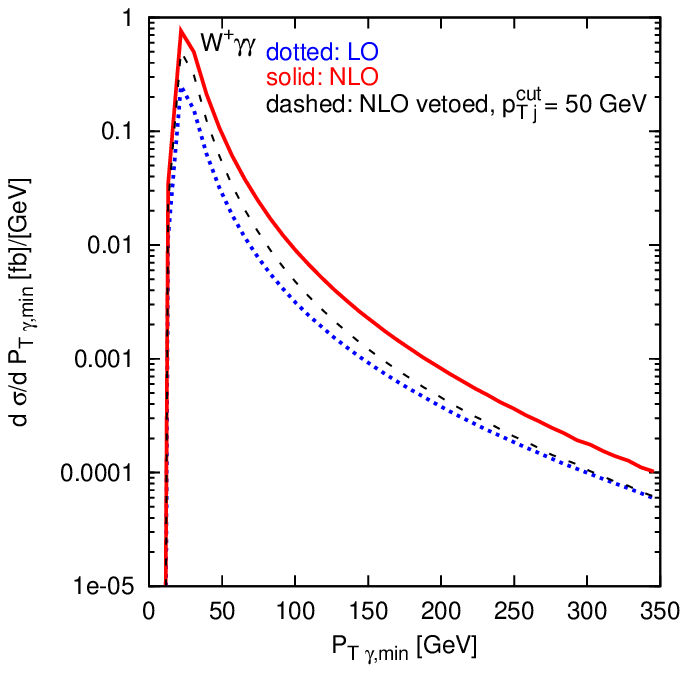}
  \includegraphics[scale = 1]{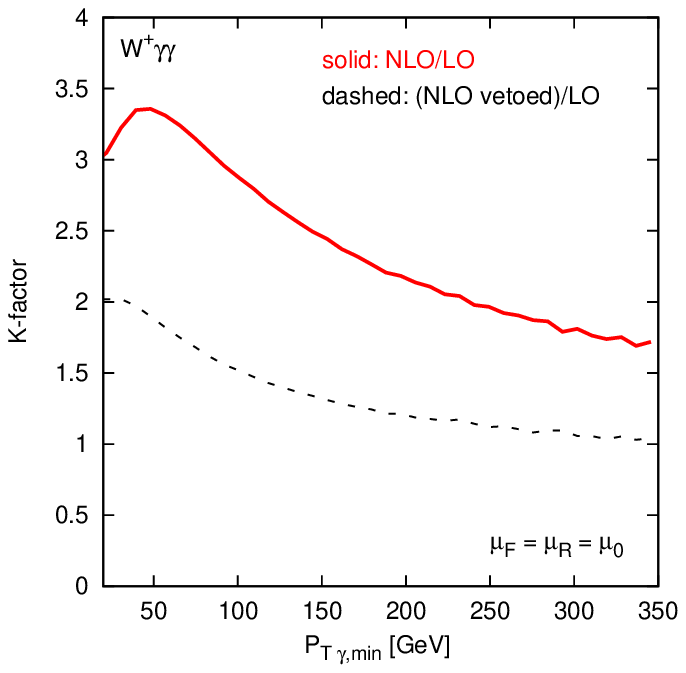}
  \caption[]{\label{fig:8}
    {\it Left:} {\sl Transverse-momentum distribution of the softest
      photon in $p p \to W^+\gamma\gamma +X \to \ell^+ \gamma \gamma
      +\sla{p}_T+X$ production at the LHC. LO and NLO results are shown for
      $\mu_F=\mu_R=\mu_0$ and the cuts of Eqs.~(\ref{eq:cuts}, \ref{eq:isol}).} 
    {\it Right:} {\sl K-factor as defined in Eq.~(\ref{eq:kfactor}).}}
\end{figure}

In Fig.~\ref{fig:6} (\ref{fig:8}), we
show the transverse momentum distribution of the harder(softer) photon
$p_{T\gamma,max}$($p_{T\gamma,min}$) in $W^+\gamma\gamma$ production. 
The K-factor is almost constant in
the case of the harder photon $p_T$ distribution. For the $p_T$
distribution of the softer photon, we get the same large K-factor
at low values of transverse momenta, but with a decrease of the NLO
enhancement above 
$p_T \approx $ 50 GeV.


\begin{figure}[tbh!]
\includegraphics[scale = 1]{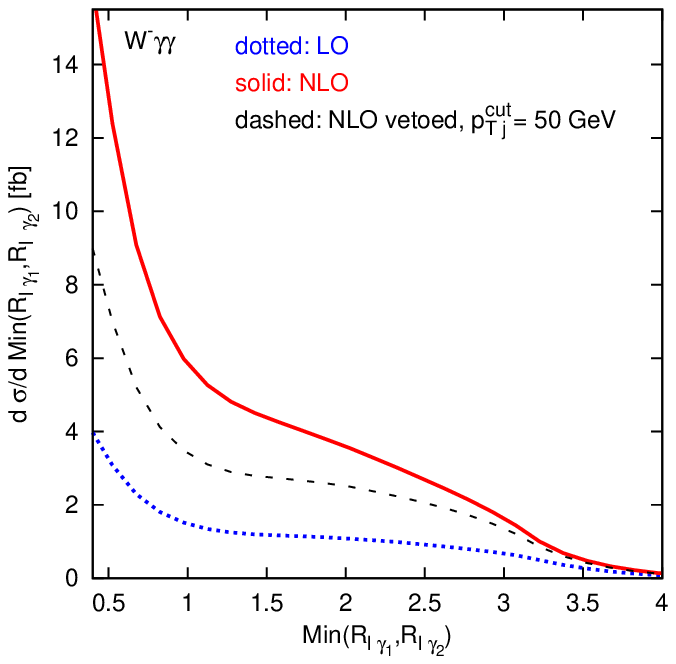}
\includegraphics[scale = 1]{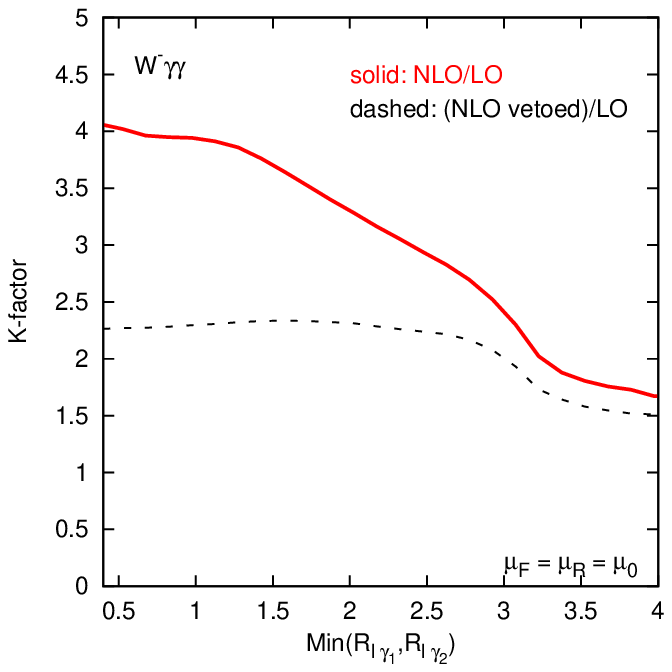}
\caption[]{\label{fig:11}
{\it Left:} {\sl Distribution of the minimum between $R_{l\gamma_1}$ and
  $R_{l \gamma_2}$ in $p p \to W^-\gamma\gamma +X \to \ell^- \gamma
  \gamma +\sla{p}_T+X$ production at the LHC. LO and NLO results are
  shown for $\mu_F=\mu_R=\mu_0$ and the cuts of Eqs.~(\ref{eq:cuts}, \ref{eq:isol}).} 
{\it Right:} {\sl K-factor as defined in Eq.~(\ref{eq:kfactor}).}}
\end{figure}

Similar features emerge for distributions of the photon-lepton
separations. In  Fig.~\ref{fig:11}, the minimum of
$R_{l\gamma_1}$ and $R_{l\gamma_2}$ is shown for $W^-\gamma\gamma$
production: the K-factor shows a large phase-space 
dependence and reaches values of 3 to 4 when the photons are
emitted close to the lepton.

\begin{figure}[tbh!]
\includegraphics[scale = 1]{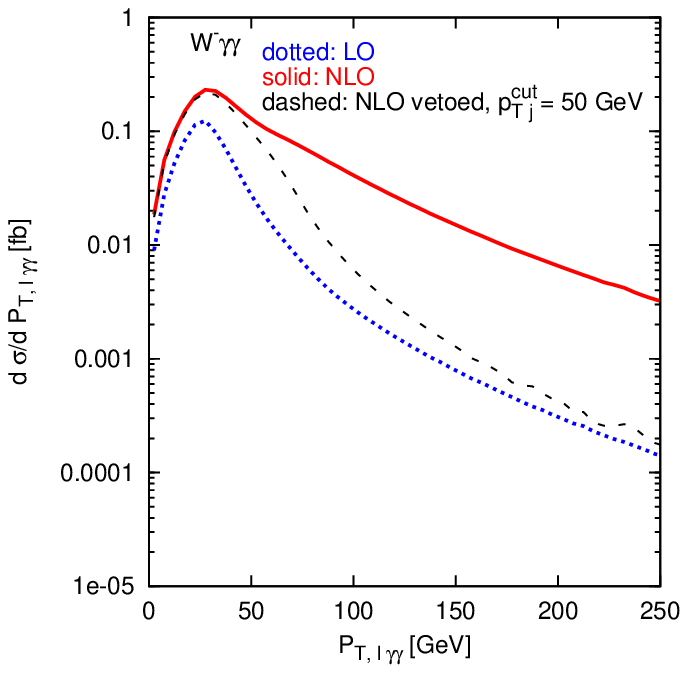}
\includegraphics[scale = 1]{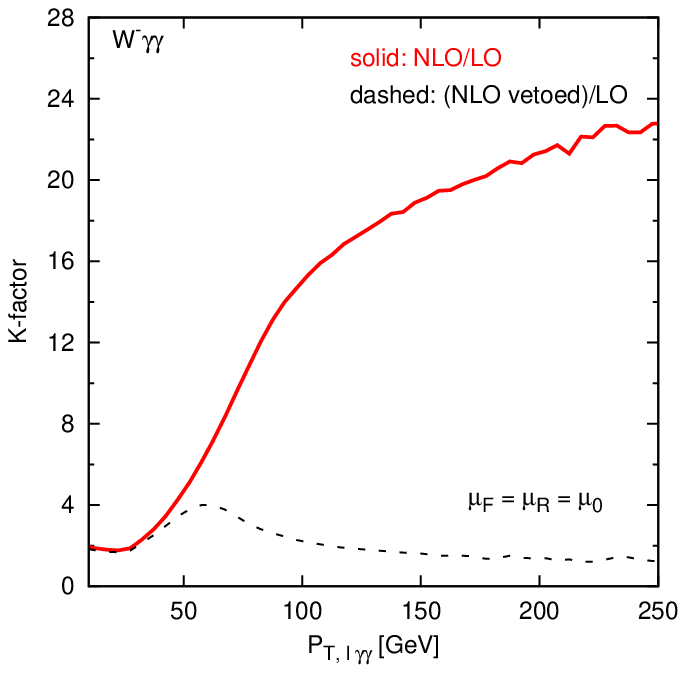}
\caption[]{\label{fig:13}
{\it Left:} {\sl Transverse-momentum distribution of the
  $\ell^-\gamma\gamma$ system in $p p \to W^-\gamma\gamma +X \to \ell^-
  \gamma \gamma +\sla{p}_T+X$ production at the LHC. LO and NLO results
  are shown for $\mu_F=\mu_R=\mu_0$ and the cuts of Eqs.~(\ref{eq:cuts}, \ref{eq:isol}).} 
{\it Right:} {\sl K-factor as defined in Eq.~(\ref{eq:kfactor}).}}
\end{figure}

Much larger values of the K-factor are possible in phase space regions
which are difficult to access due to the LO kinematics. An example with a
K-factor larger than 20 is shown in Fig.~\ref{fig:13}, where the
transverse-momentum distribution of the lepton-photon-photon system is
plotted. This large K-factor, indeed, is a feature shared by other
triple vector boson production channels: the increase of the
transverse-momentum of the observable electroweak system is compensated
by a hard jet which appears first at NLO.
\begin{figure}[t!]
\hspace*{-1cm}
\includegraphics[scale = 0.45,angle=270]{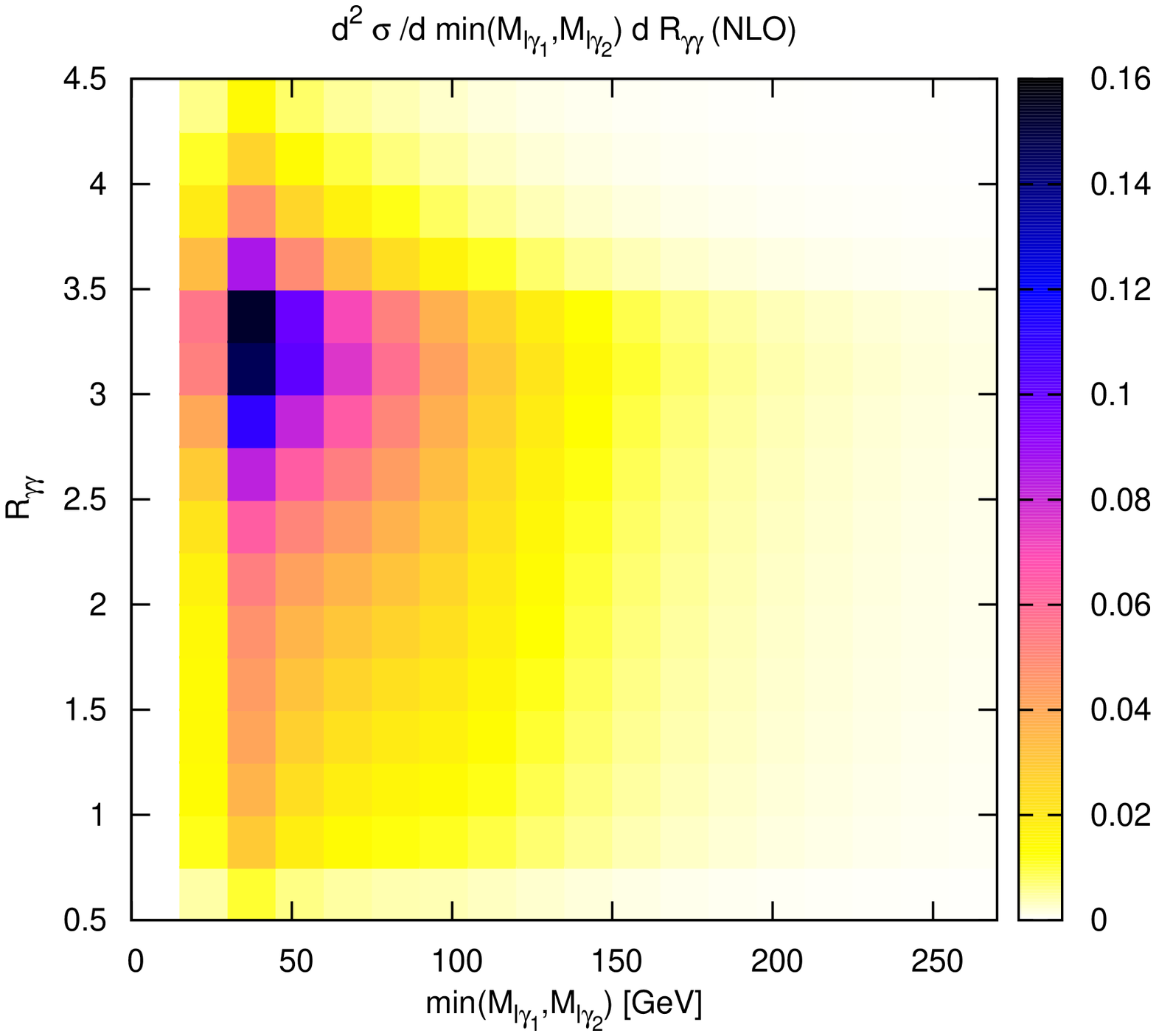}
\includegraphics[scale = 0.45,angle=270]{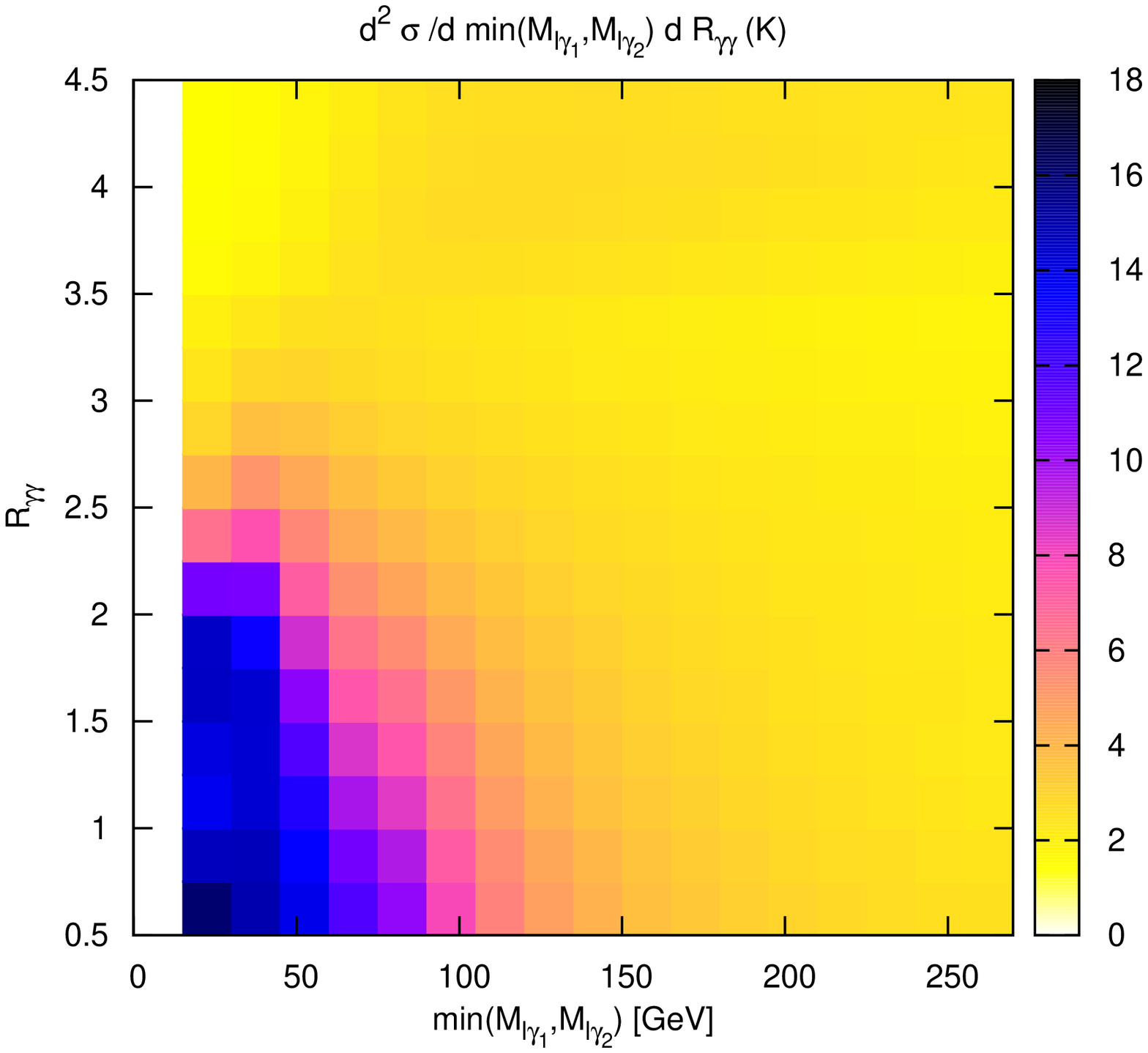}
\caption[]{\label{fig:14}
{\it Left:} {\sl Dependence of the NLO cross section on the couple of
  variables
  $(\min(M_{l\gamma_1},M_{l\gamma_2}),R_{\gamma_1\gamma_2})$. Same
  scales and cuts as in the previous diagrams.} 
{\it Right:} {\sl K-factor as defined in Eq.~(\ref{eq:kfactor}).}}
\end{figure}
\begin{figure}[tbh!]
\hspace*{-1cm}
\includegraphics[scale = 0.45,angle=270]{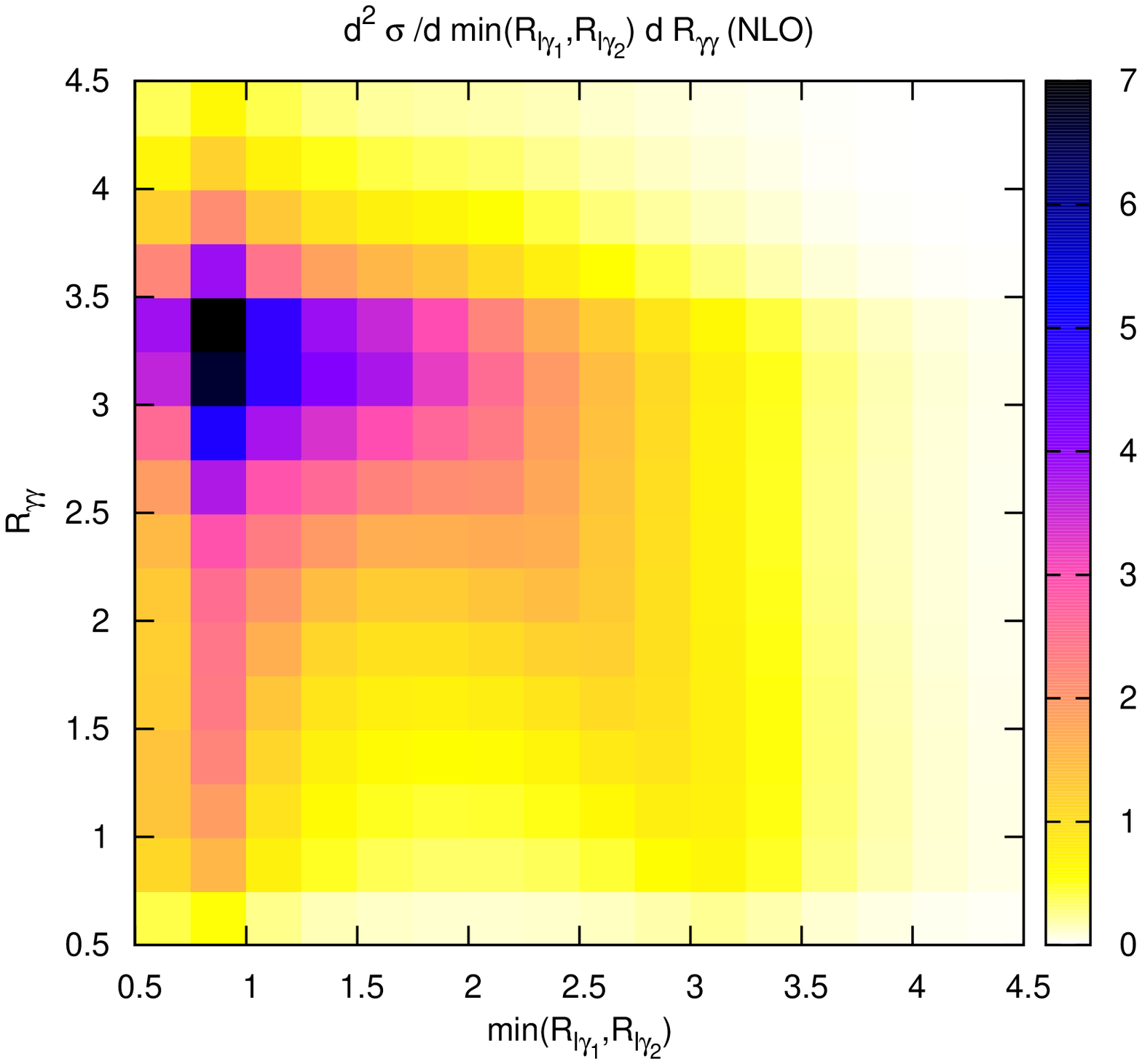}
\includegraphics[scale = 0.45,angle=270]{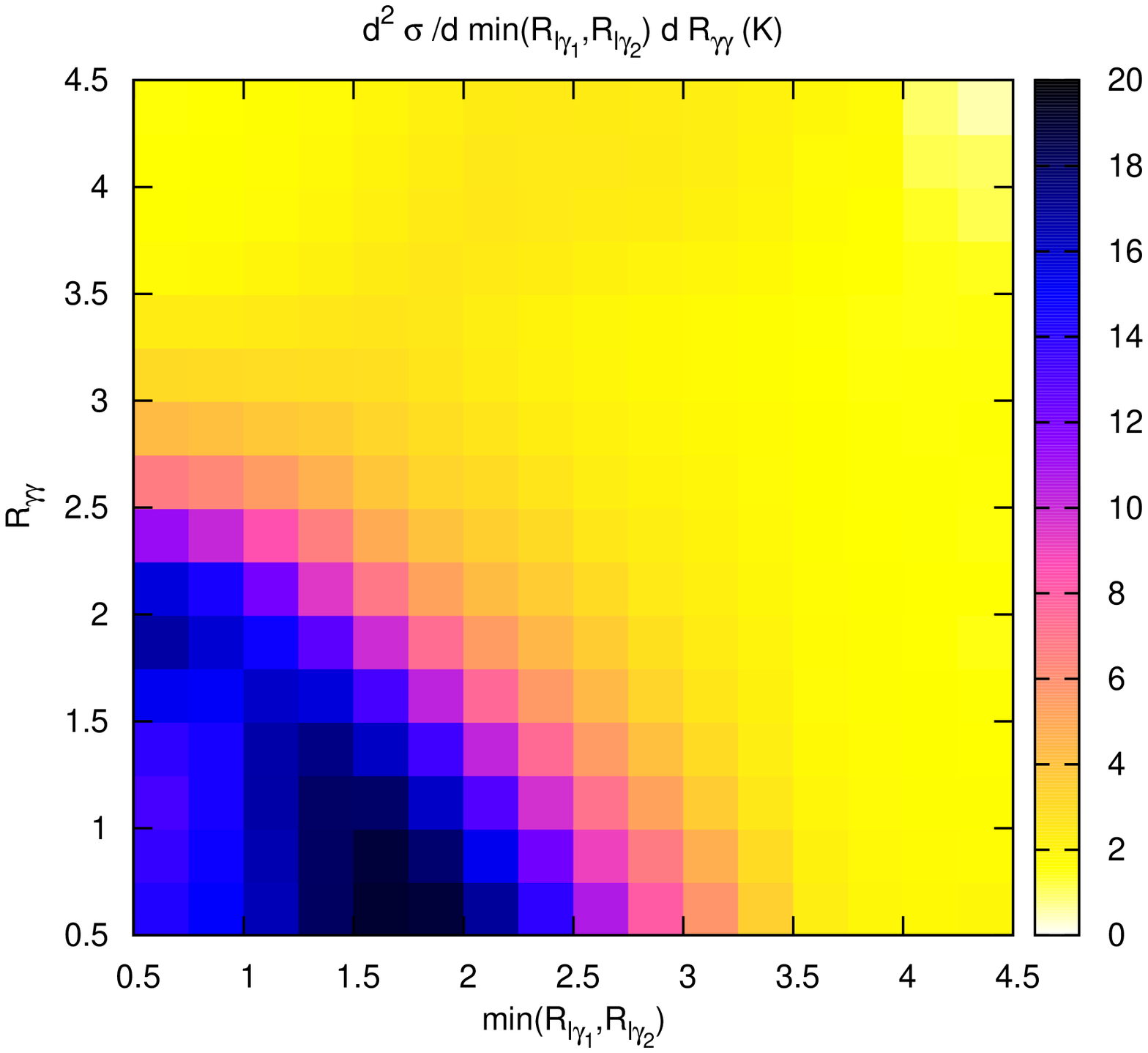}
\caption[]{\label{fig:15}
{\it Left:} {\sl Dependence of the NLO cross section on the couple of
  variables $ (\min(R_{l\gamma_1},R_{l\gamma_2}),R_{\gamma_1\gamma_2})$. Same scales and cuts
  as in the previous diagrams.}
{\it Right:} {\sl K-factor as defined in Eq.~(\ref{eq:kfactor}).}}
\end{figure}

Large K-factors also appear in the two-dimensional distributions of
Figs.~\ref{fig:14} and~\ref{fig:15}, where the dependence of the cross
section on the pairs of variables 
$(\min(M_{l\gamma_1},M_{l\gamma_2}),R_{\gamma_1\gamma_2})$ and 
$(\min(R_{l\gamma_1},R_{l\gamma_2}),R_{\gamma_1\gamma_2})$,
respectively, are shown, together with the
corresponding K-factors. It is evident that very large NLO
corrections arise particularly in regions where the lepton and photons
are emitted close to each other and for low invariant masses of the
lepton-photons system. However, since these regions contribute very little to the total
cross section, this does not affect the overall K-factor, which is similar
to the one observed in the bulk of the corrections in the differential
distributions of Figs.~\ref{fig:4} to \ref{fig:13}. Note, however, that
for the study of anomalous couplings, frequently extreme kinematics are
selected. Therefore, large K-factors might appear.
%
%

\subsection{Radiation Zero}
\label{sec:Rapidis}
As is known from the general theorem of Ref.~\cite{Brown:1982xx},  the SM
amplitude for the process  $q \bar{Q}  \to W^\pm\gamma \gamma $ vanishes
for $ \cos \theta^*_{W} = \pm 1/3$ when the two photons are
collinear. Here, $\theta^*_W$ denotes the angle between the incoming
quark and the $W$ boson in the parton-center of mass frame. This radiation
zero is not present in the gluon-induced channels, which 
enter in this process at NLO and are important at the LHC due to the steep
rise of the gluon pdfs with smaller Feynman-x. 
The radiation zero only remains present when additional neutral
(e.g. gluonic) radiation is collinear to the photon. Hence, additional
QCD emission, as part of the NLO contribution to \waa~production, is
expected to spoil the radiation zero, similar to the $W\gamma$
production process~\cite{Baur:1993ir}.
The resulting strong increase of the cross section near
$\cos\theta^*_{W} = \pm 1/3$ can explain the large
total K-factor for the \waa~process.

The radiation zero can be investigated, following
Ref.~\cite{Baur:1997bn}, via rapidity difference distributions of the W
and the photon pair system, $y_{\gamma\gamma}-y_W$, where $y_W$ is obtained
from the $W$ momenta reconstructed out of the lepton and the missing
transverse momentum. 
In our discussion we follow Ref.~\cite{Baur:2010zf} and present results only for $W^+\gamma \gamma + X$ production
since similar features are observed for $W^- \gamma \gamma + X$ production. The new element in
the present discussion is that final state photon radiation (off the
charged lepton) is included in the calculation.

\begin{figure}[h!]
\begin{minipage}[b]{1.02\linewidth}
\hspace*{-0.9cm}\includegraphics[scale =  1]{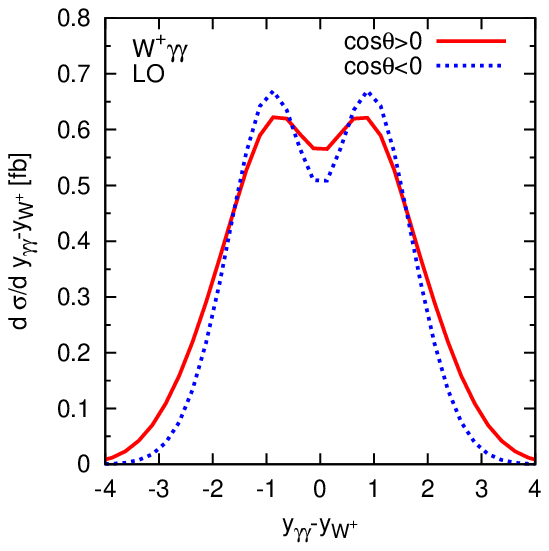}
\hspace*{-0.5cm}\includegraphics[scale = 1]{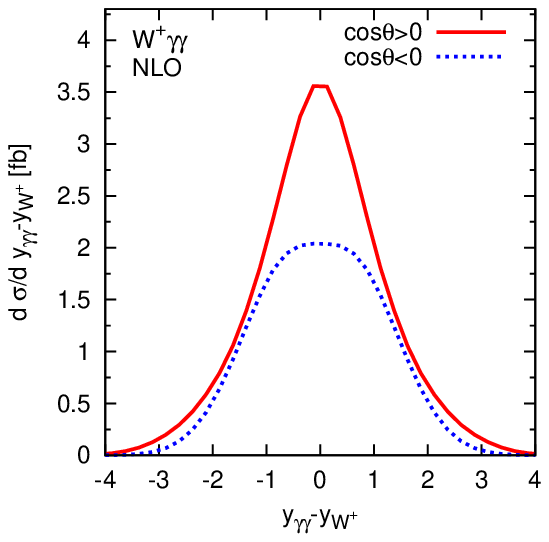}
\hspace*{-0.5cm}\includegraphics[scale = 1]{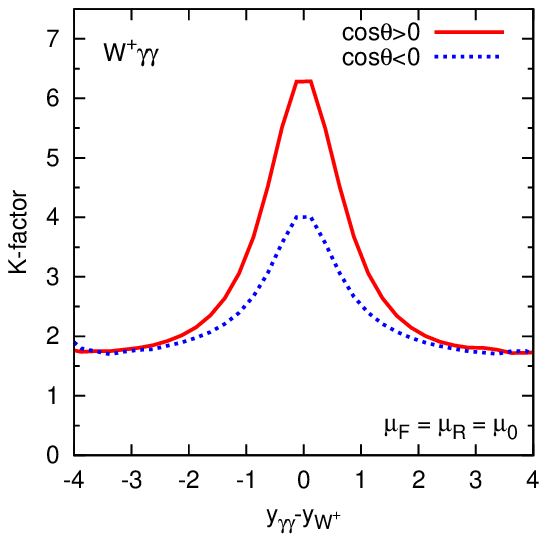}
\begin{minipage}[b]{0.98\linewidth}
\caption[]{\label{fig:16}
{\sl Distribution in rapidity separation between the W and
  the photon pair, with the photons in the same ($\cos\theta>0$) or
  opposite ($\cos\theta<0$) hemispheres, at LO (left), NLO
  (center) and K-factor (right). }}
\end{minipage}
\end{minipage}
\end{figure}

\begin{figure}[h!]
\begin{minipage}[b]{1.02\linewidth}
\hspace*{-0.9cm}\includegraphics[scale = 1]{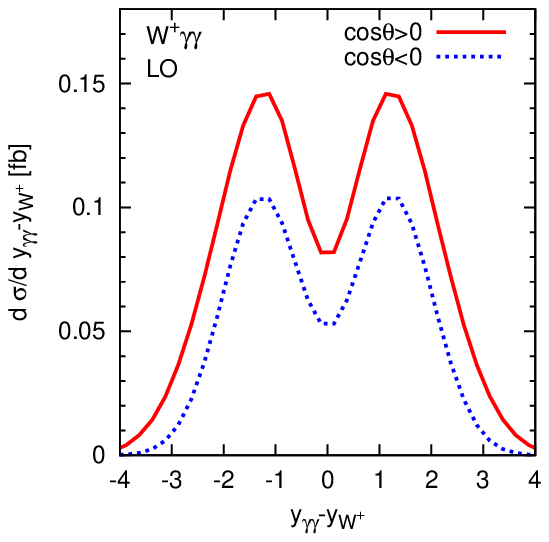}
\hspace*{-0.5cm}\includegraphics[scale = 1]{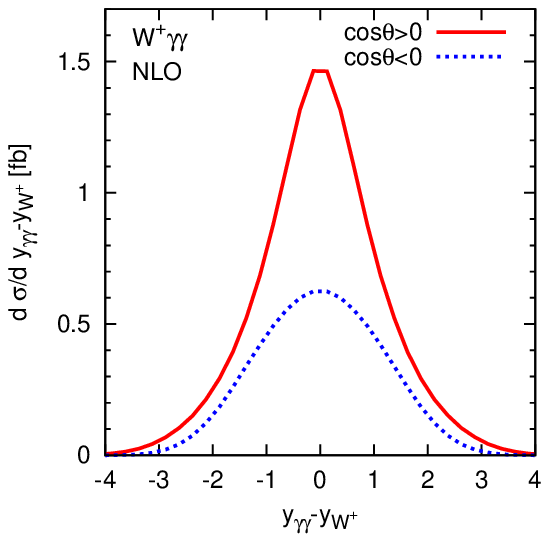}
\hspace*{-0.5cm}\includegraphics[scale = 1]{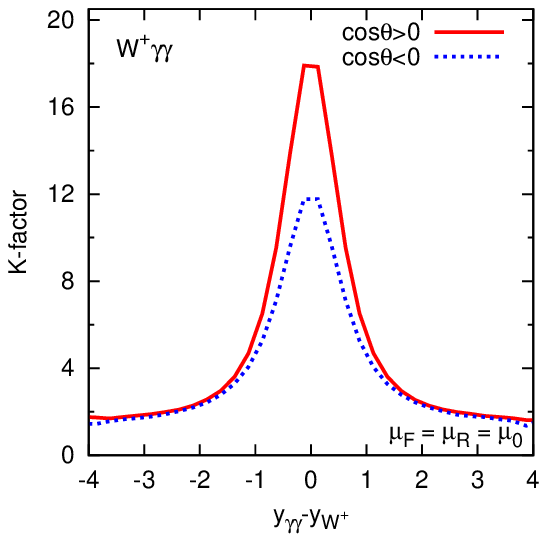}
\begin{minipage}[b]{0.98\linewidth}
\caption[]{\label{fig:17}
{\sl Same plots as in Fig.~\ref{fig:16} with the additional cut
on $M_{T,l\nu}>70$ GeV.}}
\end{minipage}
\end{minipage}
\end{figure}
In Fig.~\ref{fig:16}, we plot the distribution in rapidity separation
between the W and the photon pair, for photons lying in the same
($\cos\theta>0$) or opposite ($\cos\theta<0$) hemispheres in the
laboratory frame. Our
results at LO differ from the left-hand plot shown in Figure 3 of
Ref.~\cite{Baur:2010zf}, where the W is considered as stable: 
our zero-rapidity dip for $\cos\theta<0$ is
stronger than that for $\cos\theta>0$, while the opposite behaviour
is observed in Ref.~\cite{Baur:2010zf}. In both cases, the NLO corrections almost
fill the dips, making an observation of a radiation zero at the
LHC very difficult. In the K-factor plot of  Fig.~\ref{fig:16}(c), the
origin of the large total K-factor~($\sim$3) for \waa ~production is
clearly visible. This can be compared to other triple vector boson
production processes with typical K-factor values between 1.5 and 
2~\cite{Lazopoulos:2007ix,Hankele:2007sb,Campanario:2008yg,Binoth:2008kt,last,Bozzi:2010sj}.
The suppression in the central region at LO due to the radiation zero is
spoiled by the extra jet emission at NLO, giving rise to large
K-factors in the bulk of the corrections and therefore large total
K-factors. This is opposite to Figs.~\ref{fig:13}-\ref{fig:15}, 
where large K-factors appear in marginal phase space regions.

%
The difference observed at LO in Ref.~\cite{Baur:2010zf} as compared to our
results is due to the extra radiation of photons from
the lepton line.  In Fig.~\ref{fig:17}, we try to suppress this
contribution by restricting the phase space for photon radiation in $W$
decay. Imposing a cut, $M_{T,l\nu}>70$ GeV, on the transverse mass of
the charged lepton and neutrino, which is close to the kinematical limit
for on-shell $W$s, strongly disfavors final state photon emission.
This results in  $y_{\gamma\gamma}-y_W$ distributions with a more
pronounced dip for  $\cos\theta>0$, in agreement with
Ref.~\cite{Baur:2010zf}.  Note that even larger K-factors
now appear in the central region, reaching values close to 20 in regions
of large cross section.

\begin{figure}[tb]
\includegraphics[scale = 1]{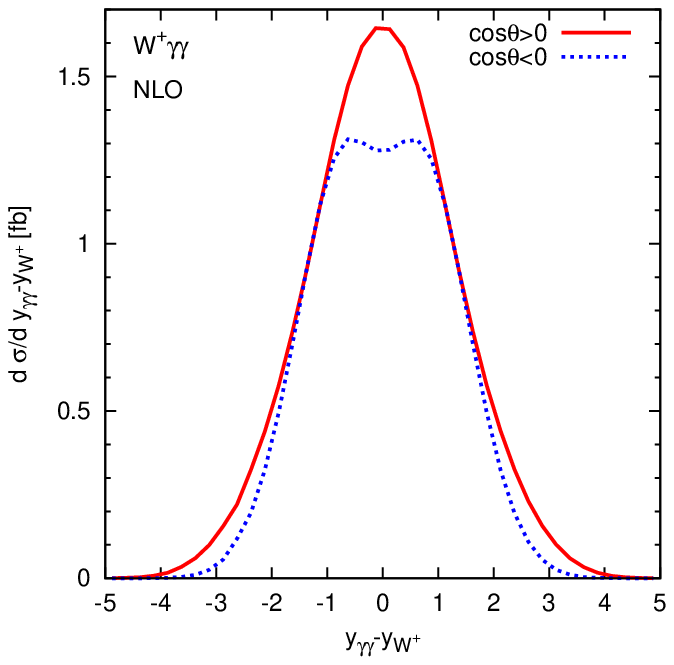}
\includegraphics[scale = 1]{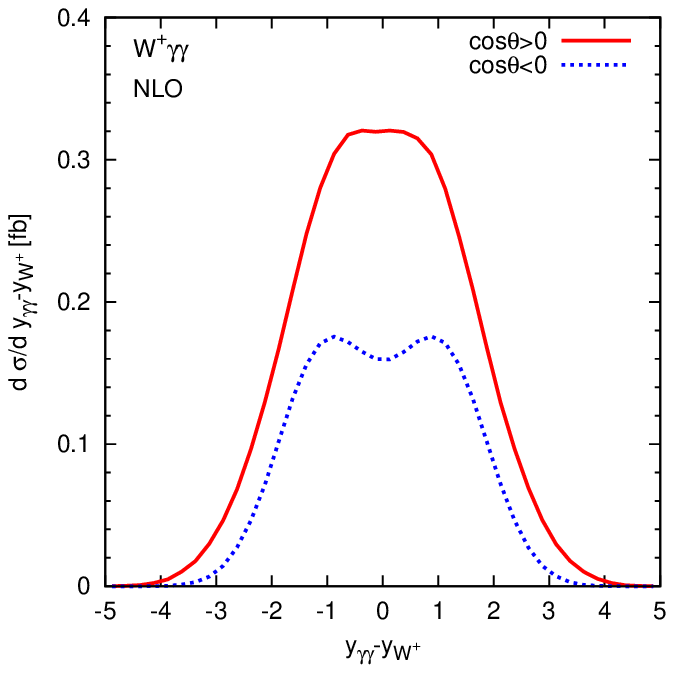}
\caption[]{\label{fig:19}
{\it Left plot: }{\sl NLO plot of Fig.~\ref{fig:16} with the
  additional jet veto cut $p_{T j}<50$ GeV.} 
{\it Right plot: }{\sl NLO plot of Fig.~\ref{fig:16} with the
  additional cuts $p_{T j}<50$ GeV and $M_{T,l \nu}>70$ GeV.}}
\end{figure}

For a better comparison with Reference~\cite{Baur:2010zf}, we also show our numerical results with
an additional jet veto cut, $p_{T j}< 50$ GeV, in Fig.~\ref{fig:19}, 
with or without the cut on the invariant mass
of the lepton-neutrino pair. Note also that,
following Ref.~\cite{Baur:1997bn}, we use true rapidity 
and not pseudorapidity distributions, as was done in
Ref.~\cite{Baur:2010zf}. We have 
checked that our pseudorapidity plots, once we unphysically remove the extra
radiation of the photon from the lepton line, are similar to those
of Ref.~\cite{Baur:2010zf}.  It is interesting to see in the last
plots, similar to Fig.~3 of Ref.~\cite{Baur:2010zf}, 
that once extra radiation is vetoed, a small dip is visible at NLO. This
feature is particularly apparent in the right figure of
Fig.~\ref{fig:19},  where final state photon
radiation is also reduced. However, whether this dip should be
interpreted as a sign of the radiation zero is questionable since it is
more pronounced for photons in opposite hemispheres. As was pointed out
in Ref.~\cite{Dobbs:2005ev}, ”signing” the quark direction
according to overall boost of the event along the beam axis  efficiently lifts
the proton-proton initial state degeneracy at the LHC. Thereby, it reduces the
distortion to the radiation zero by QCD activity and might aid in a more
significant observation of the radiation zero. We have not investigated 
these questions in the present work.

%

\section{Conclusions}
\label{sec:concl}

We have calculated the NLO QCD corrections to the processes $pp,p\bar
p\to$ \waa$+X$ with full leptonic and radiative decays of the $W$ boson. This
process can be relevant as a background for New Physics searches and as
a signal for the measurement of triple and quartic couplings at the
LHC. It also represents an irreducible background to $WH$ production, when
the Higgs boson decays into two photons.

Our numerical results show that this process gets very large corrections
at NLO, even when compared to other triple boson production processes. The
exceptionally large total K-factor is due to a cancellation between
different diagrams at LO, leading to an approximate radiation zero, 
an effect which is spoiled by the extra jet emission at NLO.
In certain regions of phase space the cross section increases by a
factor of 20 with respect to the LO calculation. Apart from the vicinity
of the radiation zero, this strong enhancement happens when the recoil
against a hard jet at NLO is kinematically favorable. This latter effect
was already observed in the NLO corrections to other triple boson
production processes. However, these high transverse momentum effects do
not significantly contribute to the bulk of the cross section
corrections and do not affect the total K-factor. Nevertheless, for the
study of 
anomalous couplings, extreme kinematics are typically selected. Therefore, large
K-factors might appear and the NLO corrections have to be taken into account.

\waa\, production at the LHC
provides an additional example of a cross section whose theoretical errors
at LO are  substantially underestimated by considering scale variations
only: the LO factorization scale variation is much smaller than the size
of NLO corrections. Remaining NLO scale variations are at the 10\% level for
the integrated \waa\, production cross section at the LHC when varying
$\mu_R=\mu_F=\mu_0$ by a factor of 2 around the reference scale 
$\mu_0=m_{WW\gamma}$. 

Given the size of the higher-order corrections and, in particular, their
strong dependence on the observable and on different phase space regions
under investigation, a fully-flexible NLO parton Monte Carlo for \waa\,
production is required to match the expected precision of the LHC
measurements. We plan to incorporate this and other processes with a
final state photon into the \vbfnlo \, package in the near future.   

\section*{Acknowledgments}

This research was supported in part by the Deutsche
Forschungsgemeinschaft via the Sonderforschungsbereich/Transregio
SFB/TR-9 ``Computational Particle Physics'' and the Initiative and
Networking Fund of the Helmholtz Association, contract HA-101(``Physics at
the Terascale''). F.C.~acknowledges partial support by European FEDER and Spanish
MICINN under grant FPA2008-02878. The Feynman diagrams in this paper
were drawn using Axodraw~\cite{axo}.

\end{document}